\documentstyle[11pt,newpasp,twoside,epsf]{article}
\def\edcomment#1{\iffalse\marginpar{\raggedright\sl#1\/}\else\relax\fi}
\reversemarginpar

\newcommand{\water}{$\mbox{H}_{2}\mbox{O}$}
\newcommand{\ammonia}{$\mbox{NH}_{3}$}
\newcommand{\formaldehyde}{$\mbox{H}_{2}\mbox{CO}$}
\newcommand{\ionhy}{H{\sc ii}}
\newcommand{\UCHII}{UCH{\sc ii}}
\newcommand{\kms}{$\mbox{km~s}^{-1}$}

\begin{document}
\title{Masers : High resolution probes of massive star formation}
 \author{S P Ellingsen}
\affil{School of Mathematics and Physics, University of Tasmania, Private
  Bag 37, Hobart 7001, TAS, Australia}

\begin{abstract}
Astrophysical masers are one of the most readily detected signposts of
high-mass star formation.  Their presence indicates special
conditions, probably indicative of a specific evolutionary phase.
Masers also represent the ultimate high-resolution probe of star
formation with the potential to reveal information on the kinematics
and physical conditions within the region at milli-arcsecond
resolution.  To date this potential has largely remained unfulfilled,
however, recent advances suggest that this will soon change.

The key to unlocking the potential of masers lies in identifying where
they fit within the star formation jigsaw puzzle.  I will review
recent high resolution observations of OH, water and methanol maser
transitions and what they reveal.  I also briefly discuss how
multi-transition observations of OH and methanol masers are being used
to constrain maser pumping models and through this estimate the
physical conditions in the masing region.
\end{abstract}

\section{Introduction}

Masers are one of the most readily observable signposts of massive
star formation and they possess characteristics that mean they have
the potential to be very powerful high-resolution probes of these
regions.  Towards massive star forming regions strong masers are
frequently observed from the main-line transitions of OH (Caswell,
Haynes \& Goss 1980), the 22 GHz transition of \water\/ (Comoretto et
al. 1990) and the 6.7 GHz (Menten 1991) and 12.2 GHz (Batrla et al.
1987) transitions of methanol.  They are more rarely observed from SiO
(see Greenhill et al. these proceedings), \formaldehyde\/ (Hoffman et
al. 2003) and \ammonia\/ (Mauersberger Wilson \& Henkel 1986) and from
other transitions of OH, \water\/ and methanol.  

The presence of maser emission indicates that the physical conditions
within the region lie within a certain range.  However, despite the
fact that the first masers were discovered towards star forming
regions nearly 40 years ago (Weaver et al. 1965) the precise physical
conditions that give rise to the various maser transitions in massive
star forming regions are still not understood.  There are two
fundamental reasons for this, the first is that the pumping of the
masers is an intrinsically complex and non-linear process.  The second
reason is that aside from information obtained directly from
observations of the maser itself (such as intensity, brightness
temperature, line width and polarization properties) there is
typically little or no complementary information available at
resolutions comparable to the size of the masers.  Towards some star
formation regions multiple different OH maser and/or methanol maser
transitions are detected.  If these transitions arise from the same
region of gas then this places significant additional constraints on
maser pumping schemes.  To date multi-transition observations of OH
(Cesaroni \& Walmsley 1991, Gray et al. 2001) and methanol (Cragg et
al. 2001, Sutton et al. 2001, Ellingsen et al. 2003) masers have only
been made towards a small number of sources.  Refinements in
observational techniques and maser modeling offer the possibility
that future multi-transitional observations may yield vital
information on the physical conditions in star-formation regions at
high-resolution.

The most common OH, water and methanol maser transitions have been
detected towards hundreds of massive star forming regions within the
Galaxy and many regions show emission from all three species.  The
exact relationship between the various maser molecules, and the
environment and evolutionary phase that they each trace is currently
understood at only a very superficial level.  Studies of the
relationship between OH and water masers has found that in regions
where these two transitions coexist they arise in nearby (separations
of less than a few arcseconds), but different locations (Forster \&
Caswell 1989).  A detailed study of methanol and water masers by
Beuther et al. (2002) found a similar relationship for these two
species.  In contrast towards many high-mass star forming regions
there is a close association between the presence and location of OH
and methanol masers and initially most methanol masers were discovered
towards known OH masers.  High resolution studies have found that in
many cases the clusters of OH and methanol masers are coincident at
the sub-arcsecond level (Caswell 1997), but individual OH and methanol
maser spots arise at differing locations (Menten et al. 1992).
However, there are many methanol and water maser sites which are not
associated with OH masers and untargetted searches for methanol masers
have detected many not associated with known star formation regions
(Ellingsen et al. 1996).  In addition, centimetre radio continuum
observations towards a large sample of methanol masers have also
failed to detect emission stronger than 1 mJy towards the majority
(Phillips et al.  1998, Walsh et al. 1998), although all are
associated with millimeter continuum sources (Beuther et al. 2002).
The generally accepted current picture is that water and methanol
maser emission commences prior to the formation of a detectable
ultra-compact \ionhy\/ (\UCHII) region, with the OH masers arising
after the commencement of fusion in the core.  There is a period
during which all three species can coexist, but relatively soon after
the formation of a detectable \UCHII\/ the methanol masers are
destroyed, followed by the water masers with the OH masers surviving
until later in the evolutionary process.  A more detailed discussion
of the relationship between the various maser species in star
formation regions is given in Caswell (2002) and Beuther et al. (2002)

\section{Masers - the ultimate high resolution probes}

As stated above, the presence of a maser within a star forming region
indicates the existence of particular physical conditions.  In most
regions where masers are detected it is likely that there are large
volumes of gas in the right physical condition to mase, but we only
observe a maser where by chance there is a long path which exhibits a
large degree of velocity coherence along our line of sight.  Under
this picture there is nothing special about our particular line of
sight and observers along different lines of sight will just detect
maser emission from different locations within the larger masing
region.  It also explains why the masers are intrinsically small.
Being both small and bright means that it is possible to image masers
using Very Long Baseline Interferometry (VLBI) and as they are
spectral lines we are able to obtain both the relative position and
light of sight velocity of the different maser spots.  It is often
possible to determine the positions of the different maser spots
within a region to sub milli-arcsecond (mas) accuracy relative to a
chosen reference maser spot or to some nearby background continuum
reference source.  To put this in perspective, for a massive star
formation region at a distance of 2~kpc an angular resolution of 1 mas
corresponds to a linear resolution of 0.3 AU.  The small size of the
masers and the high accuracy to which their relative positions can be
determined means that it is possible to measure proper motions of the
masers on time-scales as short as weeks, but more typically on
time-scales of years (e.g. Genzel et al. 1981a).  This makes masers
very powerful probes of the kinematics in the regions where they
exist.  To date, it is as probes of the kinematics that masers have
made the most striking contribution to our understanding of massive
star formation at high-angular resolution and section 4 discusses some
specific examples of this.

Masers are also able to yield information on the magnetic fields
within the regions where they form.  The OH molecule is paramagnetic
and this means that it undergoes significant Zeeman splitting in the
presence of interstellar magnetic fields.  The Zeeman splitting causes
OH masers to be highly circularly polarized, sometimes up to 100\%.
VLBI observations can be used to identify Zeeman pairs (left and
right-hand circularly polarized OH masers with different velocities
that are coincident at the milli-arcsecond level) and the total
magnetic field strength in the masing region can then be determine
from the velocity difference between the two orthogonal polarizations
(see for example Caswell (2002)).  The size of the velocity separation
for the Zeeman pairs depends upon the Lande-g factor for the
transition and differs for each OH maser transition.  The excited OH
transitions at 6030 and 6035 MHz have both smaller Zeeman splitting
and generally simpler spectra than the main-line transitions which
means that it is possible to confidently identify orthogonal
polarization pairs from single dish spectra alone, particularly when
observations are made of more than one transition (Caswell 2003).

\section{Masers - the Bart Simpson of star formation research}

If as outlined in the previous section masers are such powerful probes,
then it is reasonable to ask why they are not more prominent in the
field of star formation research?  Studies of masers play a prominent
role in studies of AGB and post-AGB stars and maser observations have
been the key to a number of new and exciting discoveries relating to
active galaxies.  So why is it that for star forming region, masers
(like Bart Simpson) are under-achievers?  I believe that the most
important factor is that in contrast to AGB stars and active galaxies
there is relatively little complementary high-resolution information
at other wavelengths available for star formation regions.  The
earliest stages of high-mass star formation take place deep within
cold, dense clouds of gas and dust which are often optically thick at
wavelengths as long as 20~$\mu$m.  In addition, high-mass stars form
in clusters and so high resolution observations are essential to
disentangle the different phenomena associated with stars of a
variety of masses at a various evolutionary phases.  High resolution
observations in the mid and far-infrared and sub-millimeter regimes
are only now becoming a reality and these are set to revolutionize our
understanding of massive star formation.  Far-infrared and
sub-millimeter images of high-mass star formation at arcsecond
resolution will help to resolve the confusion of the cluster
environment and characterise the region in which the masers arise.  I
believe that a better understanding of the environment in which the
masers form will unlock their potential as probes of high-mass star
formation, much as it has for AGB stars and active galaxies.

\section{Masers as high resolution probes}

There are a number of sources where studies of the proper motion of
masers in high-mass star formation regions have made a significant
contribution to our understanding of the kinematics.  The rest of this
paper will detail some of the best examples of masers as kinematic
probes.

\subsection{Water masers in Cepheus A}

In the late 1970's and early 1980's the first VLBI proper motion
observations of water masers were undertaken towards sources like
Orion KL and W51 (Genzel et al. 1981a,b).  A relatively common feature
of water maser spectra in massive star forming regions is strong
emission at the systemic velocity and weaker emission at velocities
offset by $\pm$50-200 \kms .  The observations of Genzel et al.
revealed that the systemic velocity features typically show large
proper motions, while the high-velocity features show much smaller
proper motion.  This is readily explained if the water masers arise in
regions of post-shocked gas.  If the direction of propagation of the
shock is close to the plane of the sky then the line of sight velocity
will be approximately the systemic velocity, but the transverse
velocity and hence the proper motion will be large.  Conversely if the
shock is propagating in a direction close to the line of sight then
the velocity of the maser emission will be offset from the systemic
velocity of the region, but the proper motion will be small.  The path
length through the post-shocked gas is also much smaller in this case
which explains why the high-velocity maser features are weaker than
those at the systemic velocity.  Proper motion experiments enable the
three-dimensional velocity distribution of shocks and outflows in
high-mass star formation regions to be traced.  However, the maps of
maser water maser emission and proper motions of Genzel et al. and
others show that in practice the overall distribution of masers is
complex and since location of protostars, YSO etc. in the regions is
typically unknown, interpretation of the maser proper motions is
difficult.

There are a small number of water maser sources where it has been
hypothesised that the emission arises in a circumstellar disk.
Finding unambiguous evidence for maser emission in a disk (the
equivalent of NGC4258 for active galaxies), is something of a holy
grail for researchers studying masers in star formation regions.
There is good evidence for the existence of disks around high-mass
stars from infra-red and thermal molecular line observations (see for
example papers by Padgett and Wilner in these proceedings).  Masers
offer the possibility of revealing detailed information on the disk
dynamics which could be used to infer disk dimensions, whether it is
warped by instabilities, accretion rates etc.  To date none of the
purported disks in star-forming regions are as convincing as that
observed towards the nucleus of NGC4258 (Miyoshi et al. 1995), but
proper motion observations of a possible disk associated with Cepheus
A HW2 by Torrelles et al. (2001) have discovered an even more
intriguing object.
  
The tale of Cepheus A HW2 is a theme common to high resolution
observations of masers from a variety of transitions.  Observations
with arcsecond resolution with a connected element interferometer (in
this case the VLA) showed a flatten structure perpendicular to a radio
continuum jet with a velocity gradient and this was hypothesised to be
a disk (Torrelles et al. 1996).  To test the hypothesis further
observations using a VLBI array with milli-arcsecond resolution (in
this case the VLBA) were made, however, rather than confirming the
original hypothesis the higher resolution observations instead
revealed greater complexity.  Torrelles et al. found that what
appeared to be single maser spots at arcsecond resolution were in many
cases very narrow linear and curved structures, often showing proper
motion perpendicular to the major axis.  One of the maser spots
slightly resolved at arcsecond resolution is revealed to be a
remarkable arc of water masers.  The arc of masers extends over 0.1
arcseconds (72 AU at a distance of 725 pc), fits a circle with an
accuracy of 1 part in 1000 and is expanding with a velocity of
9~\kms .  The radius of the circle is 62 AU and it has a dynamical age of
33 years (Torrelles et al.  2001).  Recent observations with the VLA
have detected a weak radio continuum source coincident with the centre
of the expanding circle of masers, although it has not yet been
possible to identify the nature of this source (Curiel et al. 2002).
The highly circular appearance of the arc of masers is strongly
suggestive of spherical expansion and the masers are thought to be due
to either spherical episodic ejection from a protostar or; the initial
expansion of an \ionhy\/ region from a B3 or B4 star.  Either way further
study of this object seems set to yield new information on the
formation of high-mass stars and their interaction with their natal
environment.

\subsection{Methanol masers in NGC7538}

One advantage of observations of class II methanol masers such as the
6.7 and 12.2~GHz transitions is that unlike OH and water masers that
are associated with a range of astrophysical environments, methanol
masers are only found towards high-mass star formation regions (Minier
et al. 2003).  Early arcsecond resolution images of class II methanol
masers found that many of them exhibited a simple linear or curved
spatial morphology, often with a monotonic velocity gradient along the
line.  The most popular hypothesis for this observation was that the
methanol masers traced an edge-on circumstellar disk (Norris et al
1993, 1998).  However, as with the water masers in Cepheus A, the
milli-arcsecond resolution images while consistent with the arcsecond
resolution images frequently reveal more complex structures which are
not easily explained as edge-on disks (e.g. Minier, Booth \& Conway
2000, De Buizer et al. 2002).  These and other observations and
arguments suggest that the vast majority of methanol masers do not
arise in edge on disks, but are more likely to be associated with
shocks (Lee et al.  2002).  Although clearly lower velocity and less
energetic shocks than those that produce water masers.
  
Despite this there are some sources for which the milli-arcsecond
resolution images are most easily explained as an edge-on disk.  The
best example is that of NGC7538 which has a shows a single very narrow
emission feature which in the 12.2 GHz transition is approximately 45
mas (110 AU) long with a monotonic velocity gradient along it (Minier,
Booth \& Conway 1998).  Assuming Keplarian rotation, a measurement of
velocity gradient along still has a dependency between the binding
mass and the radius of the emission.  If a plausible high-mass
protostellar mass is assumed for NGC7538 (10 M$\sun$) then the implied radius of
the disk is of the order of 940 AU (Minier et al. 2000).  This implies that
either the methanol maser emission traces only a very small fraction
of the total disk or that the disk is not that of a high-mass
protostar.  In general low implied masses or small disk fractions are
a plausibility issue for many potential methanol maser disk sources
(Minier et al.  2000).

\subsection{OH masers in W3(OH)}

Like water masers, OH masers are associated with a range of
astrophysical objects, including evolved stars, SNR/molecular cloud
interfaces and high-mass star formation regions.  OH masers were first
discovered towards star formation regions and the best studied region
is W3(OH) which exhibits OH maser emission from a wide variety of
transitions, some of which haven't been observed towards any other
sources (Cesaroni \& Walmsley 1991).  The OH masers (and the class II
methanol masers) lie in a number of clusters projected in front and
just beyond the edge of an ultracompact HII region (Reid et al. 1980,
Menten et al. 1992).  As with many/most OH masers there is little or
no apparent internal structure to the clusters.  Proper motion
observations of the OH masers by Bloemhof, Reid \& Moran (1992) find
the masers to be in slowly expanding (3 \kms) gas.  This observation
supports the generally accepted picture that the OH masers arise at
the interface between the ionized and molecular gas.
  
An exciting recent development is the detection by Argon, Reid \&
Menten (2003) of weak OH maser emission associated with the
Turner-Welsh (TW) object near W3(OH). The TW object is a high-mass
protostar offset approximately 6 arcseconds from the W3(OH) \UCHII\/
region which has associated water masers and synchrotron jets.  The OH
maser emission arises at the end of the water maser emission and may
indicate that OH masers found towards star formation regions which are
not directly associated with radio continuum emission, may trace
outflows. They also offer the possibility of probing the jets from the
TW object and other similar objects at high-resolution in more than
one regime (as the OH maser emission appears to arise further out in
the jet than the water masers).

\section{Conclusions}

For a small, but growing number of sources observations of masers are
enabling high-mass star formation regions to be studied at
sub-arcsecond resolutions.  The examples above demonstrate that VLBI
proper motion studies are a powerful method for probing the kinematics
within the star forming regions, the challenge we currently face is to
understand exactly what part of the star formation process the masers
are associated with.  Multi-transition studies of OH and methanol
masers, are currently in their infancy, but offer the prospect of
being able to tightly constrain the physical conditions which produce
the masers and which will greatly enhance the information obtained from
kinematic studies.  Combined with complementary high-resolution
studies of high-mass star formation regions in the millimeter through
to mid-infrared wavelength ranges there appear to be good prospects
that in the near future masers will play a much more prominent role in
studies of high-mass star formation than is presently the case.

\section*{Acknowledgements}

This research has made use of NASA's Astrophysics Data System.

\end{document}